\documentclass[12pt]{article}

\usepackage{verbatim}
\usepackage{amssymb}
\usepackage{amsmath}
\usepackage{subcaption}
\usepackage{epsfig}
\usepackage{psfrag}
\usepackage{graphicx}
\usepackage{ulem}
\usepackage{url}
\usepackage{multirow}
\usepackage{authblk}
\usepackage{color}
\usepackage{hyperref}
\usepackage[strings]{underscore}
\usepackage{tabularx}
\providecommand{\keywords}[1]{\textbf{\textit{Index terms---}} #1}

\DeclareGraphicsExtensions{.pdf,.jpg,.jpeg,.png,.eps}
\graphicspath{../figures/}

\begin{document}

\title{A semi-supervised approach to message stance classification}

\author[1]{Georgios Giasemidis}
\author[1]{Nikolaos Kaplis}
\author[2]{Ioannis Agrafiotis}
\author[3]{Jason R. C. Nurse}

\affil[1]{\{georgios, nikos\}@countinglab.co.uk, CountingLab Ltd., Reading, UK}
\affil[2]{ioannis.agrafiotis@cs.ox.ac.uk, Department of Computer Science, University of Oxford, Oxford, UK}
\affil[3]{j.r.c.nurse@kent.ac.uk, School of Computing, University of Kent, Canterbury, UK}

\maketitle


\begin{abstract}
	Social media communications are becoming increasingly prevalent; some useful, some false, whether unwittingly or maliciously. 
	An increasing number of rumours daily flood the social networks. 
	Determining their veracity in an autonomous way is a very active and challenging field of research, with a variety of methods proposed. 
	However, most of the models rely on determining the constituent messages' stance towards the rumour, a feature known as the ``wisdom of the crowd''.  
	Although several supervised machine-learning approaches have been proposed to tackle the message stance classification problem, these have numerous shortcomings. 
	In this paper we argue that semi-supervised learning is more effective than supervised models and use two graph-based methods to demonstrate it. 
	This is not only in terms of classification accuracy, but equally important, in terms of speed and scalability.
	We use the Label Propagation and Label Spreading algorithms and run experiments on a dataset of 72 rumours and hundreds of thousands messages collected from Twitter.
	We compare our results on two available datasets to the state-of-the-art to demonstrate our algorithms' performance regarding accuracy, speed and scalability for real-time applications.
\end{abstract}

\section{Introduction}
\label{sec:introduction}

Online content is at the centre of today's information world. A primary 
source of this content is social media, with the public acting as a major contributor 
on everything from election discussions to reports on ongoing crisis events. This level of
free engagement has several benefits. For instance, it can encourage healthy discourse on 
pertinent topics of public interest, or it can be invaluable at supporting official responders 
reacting to an unfolding crisis -- such as Hurricane Harvey in the US~\cite{twsj2017} or the 
Manchester bombings in the UK~\cite{mtv2017}. On the other hand social media can be used
as a tool to disrupt and harm society. Over the last few years, we have seen a spate of 
misinformation and fake news intended to misguide, confuse and potentially 
even risk people's lives~\cite{bbcfuture2017}. This emerging reality highlights the power of social media and the need to reliably discern genuine and useful from harmful information and noise.

There has been a wide range of research in the social media domain. Of most relevance to this work is the technical effort aimed at understanding and mitigating 
any disruptive impacts (e.g. malicious rumour propagation). Such work can be found as early as in Castillo et al.~\cite{castillo2011information} where a series of automated 
methods are used to analyse the credibility of information on Twitter. The major contribution of that article 
has been the identification of an area of research aimed at automatically estimating the validity of 
rumours and features of credible content, as information spreads across social media platforms. 
Since then, there have been a number of proposals exploring information trust, credibility and 
decision-making, using technical (e.g. machine learning) and user-centred (e.g. focused on 
perceptions and behaviours)
approaches~\cite{morris2012tweeting,castillo2013predicting,gupta2013faking,nurse2014two,alrubaian2016credibility}.
These approaches may consider individual or aggregated content (posts, messages, etc.) within 
rumours and use this as a basis for credibility or trust decisions.

Most of the research on social media rumours focuses on determining their veracity. Several authors have proposed different supervised systems using temporal, structural, linguistic, network and user-oriented features~\cite{Giasemidis:2016dtv,mendoza2010twitter,vosoughi2015automatic,wu2015false}. However, these approaches assume that message annotation\footnote{Message annotation refers to the classification of the message stance towards the rumour.} is granted. Being able to annotate messages automatically is the most important step towards determining the veracity of rumours~\cite{zhao2015enquiring}. 

Therefore, one of the most intriguing areas of research in the domain of social media is the 
problem of message (i.e. post, tweet, etc.) stance classification. Here, the aim is to determine 
whether a particular message supports, refutes or is neutral towards a rumour; neutral stances
can be further expanded to differentiate between querying or commenting messages, as highlighted 
in Zubiaga et al.~\cite{zubiaga2017detection}. Stance classification is essential for the modelling of veracity in a dataset.

In this paper, we aim to improve on the current state-of-the-art by proposing a semi-supervised approach to the problem of message stance classification. We use two graph-based semi-supervised algorithms with a variety of experimental settings. We demonstrate the performance of the models on two publicly available datasets.

The novel aspects of this work are twofold. First, we propose a new machine-learning approach, based on semi-supervised learning, to the problem of message stance classification. We argue that this is a more well-rounded way to tackle the problem than using supervised learning both in terms of accuracy and, perhaps more importantly when dealing with large and diverse datasets, in terms of computational speed and scalability. We should clarify that we do not introduce a new algorithm, but we apply an existing class of algorithms to the problem for the first time.
Second, we use a larger and more diverse dataset of rumours in terms of size and topics. Our dataset consists of 15 distinct events in comparison to the publicly available ones which contain either a single event or nine events, see next sections for further details. Particularly, the lack of diversity in rumours in the publicly available datasets introduces bias/over-fitting and does not facilitate transference of knowledge, forcing the need for constant re-training.

The performance of the semi-supervised models with different features and parameters are tested on data from an earlier study~\cite{Giasemidis:2016dtv} consisting of tweets which have been manually annotated. Our work has concluded in a semi-supervised model that consists 
of the Label Spreading algorithm using 1,000 Brown Clusters (i.e. groups of words that 
are assumed to be semantically related) as features. The model's performance is 
enhanced by manually annotating a small portion of the tweets. To validate our model, we apply it to two datasets; the first consists of seven rumours from the UK riots in 2011 \cite{Lukasik2015ctl}, achieving
an $84.9\%$ accuracy while outperforming all benchmark and random models. The second set (the PHEME dataset) consists of 23 rumours from 9 major events \cite{zubiagaSemEval2016}, has a higher bias and scores $75\%$ accuracy and outperforming all other models in terms of weighted accuracy.

The remainder of this paper is structured as follows. In Section~\ref{sec:litreview} we review the literature for the state-of-art techniques in message stance classification. In Section~\ref{sec:methodology} we introduce the methodology and elaborate on the semi-supervised algorithms used in this study. Section~\ref{sec:results} presents the results from the experiments we performed using different feature sets, algorithms and kernels (Section~\ref{sec:results-label-propagation}). Furthermore, we validate the methods on two independent sets of rumours, one from the London riots and the PHEME dataset, and compare the results to the literature (Section~\ref{sec:comparison}). Finally, Section~\ref{sec:conclusion} concludes the paper and discusses future work.

\section{Related Work}
\label{sec:litreview}

The area of rumour stance classification has recently attracted the interest of the academic community. Unlike the case of rumour veracity classification where a rumour is classified as true or false, the focus of the rumour stance classification is on individual messages. More specifically, the aim is to classify messages which contribute to a rumour into four categories, namely supporting, denying, querying and commenting. It is worth mentioning that often querying and commenting classes are either omitted or merged. Thus far, most works in this area adopt supervised methods and differ mainly in the machine learning approaches used for the classification and in the set of features that are utilised in the aforementioned algorithms~\cite{zubiaga2017detection}.

The first study to delve into classification of tweets was by Mendoza et al.~\cite{mendoza2010twitter}, where a collection of rumours, whose veracity was identified, was further analysed manually to establish the number of tweets that were supporting or denying the rumour. The authors classified the tweets into those denying, confirming or questioning the rumour and the end goal was to understand if the distribution of these classes can be indicative of the veracity of a rumour. Their results suggested that for rumours whose veracity was true, 95\% of tweets confirmed the rumour. On the contrary, when the veracity of the rumour was deemed as false only 38\% of the tweets supported the rumour. Procter et al.~\cite{procter2013reading} derived similar conclusions when analysing rumours during the UK riots in 2011. They focused particularly on the popularity of the users tweeting rumours, compared patterns of how false and true rumours start and evolve and identified significant differences. Extending the afforementioned works, Andrews et al.~\cite{andrews2016keeping}, narrowed their focus on how ``official'' accounts can help contain a false rumour and offer best social media strategies for large organisations.

Qazvinian et al.~\cite{qazvinian2011rumor}, were the first to automatically classify the stance of tweets. The authors opted for Bayesian classifiers and used the same feature set that they extracted to determine the veracity of rumours. They limited their approach by considering only two classes for annotating tweets (denying and confirming). In addition, they considered only long-term rumours and focused on how users' beliefs change over this long period. In a similar vein, Hamidian et al. \cite{hamidian2016rumor}, focused on features related to time, semantic content and emoticons and their approach outperformed Qazvinian et al. They extended their previous work by introducing the Tweet Latent Vector approach and by considering what they coined as ``belief features'', which are features that investigate the level of committed belief for each tweet~\cite{hamidian2016rumor}. 

In a similar vein, Mohammad et al.~\cite{mohammad2017stance}, propose a detection system able to determine a stance of a tweet for a particular target (i.e., person, institution, event etc.) by exploring correlations between stance and sentiment. Their system draws features from word and character n-grams, sentiment lexicons and word-embedded characteristics from unlabelled data. A linear-kernel SVM classifier is utilised to produce three clusters (positive, negative and not-determined stance) with very promising results (70\% F-score on SemEval-2016 
data). We note that training for stance is not generalised across all tweets, but is restricted per target group.

Based on Werner's et al.~\cite{werner2015committed} belief tagger, Hamidian et al. \cite{hamidian2015rumor} created a vector indicating whether a user strongly believes in the proposition; provides a non-committed comment; reflects a weak belief in the proposition; does not expressing a belief in the proposition. Lexical features were also used based on bag-of-word sets, which consist of word unigrams. The authors then explored the performance of a set of classifiers, inter alia J48 Decision Trees, Naive Bayes networks and reported that Sequential Minimal Optimization (SMO) outperforms all approaches. Similarly to previously presented work, their approach is limited to long-term rumours only.

Zeng et al.~\cite{zeng2016unconfirmed} focus more on semantic and linguistic characteristics and they introduce Linguistic Inquiry and Word Count (LIWC) features, as well as n-grams and part-of-speech components. Based on their experimentation and coded dataset, they are able to achieve an accuracy of over 88\% in classifying rumour stances in crisis-related posts; here, random forest models result in the best performance. Lukasik et al.,~\cite{lukasik2016using,lukasik2015classifying} designed a novel approach based on Gaussian Processes. They explored its effectiveness on two datasets with varying distributions of stances. The authors report results on cases where all tweets encompassing a specific rumour are used for testing and cases where the first few tweets are added to the training set. The classifier performs very well in the latter case.  The novelty of this work lies in the classification of unseen rumours since this approach can annotate tweets for each rumour separately, enabling the classification of tweets for emerging rumours in the context of fast-paced, breaking news situations.

Jin et al.~\cite{jin2016news}, suggest an unsupervised topic model method to detect conflicting tweets which discuss the same topic, as a first step for determining the veracity of fake news. They determine the stance of a tweet by focusing on a pair of values (topic and view point) represented by a probability distribution over a number of tweets. The topic-viewpoint pairs are then clustered into conflicting  viewpoints when the distance between topics-viewpoints of the same topic exceeds a predefined threshold. Once conflicting tweets are determined, a graph of the network containing tweets which refer to the same topic is created and an effective loss function is used to solve the optimisation problem.

Zubiaga et al.~\cite{zubiaga2016stance}, introduce a novel approach that considers the sequence of replies in conversation threads in Twitter. Users' replies to one another were converted to nested tree forms and tweets were analysed not only based on  their individual characteristics (content, semantics etc.) but also on their position in the conversation.  Two sequential classifiers namely Linear Conditional Random Fields (CRFs) and Tree-CRFs were adopted and eight datasets were used for validation with Tree-CRF performing slightly better than the Linear-CRF. 

Kochkina et al.~\cite{kochkina2017turing}, proposed a  deep-learning approach adopting Long Short-Term Memory networks (LSTMs) for sequential classification. They perform a pre-processing step by removing non-alphabetic characters and they tokenise the words. They further extract word vectors based on Google's \textit{word2vec} model~\cite{mikolov2013efficient}, count negation words and punctuation, identify the presence of attachments, follow the relation of content to other tweets in the discussion and count the content length. The model is trained using the categorical cross entropy loss function, however, they report that their approach is unable to distinguish any tweets denying a rumour, which are the most under-represented in their dataset. They note that these tweets are mostly misclassified as commenting and theorise that an increased amount of labelled data would improve the performance of their approach.

It is also worth mentioning some approaches that have critically reflected on the literature of stance categorisation. Shu et al.~\cite{shu2017fake}, present an overview of emerging research regarding fake news and stance classification. They elicit features from psychology and social theories, linguistic examination, as well as network and user characteristics and identify a number of models that can potentially utilise such features. One of these is stance-based approaches which centre around a single post and propagation-based approaches which focus on how tweets about a theme are interconnected. They conclude their survey by proposing a number of different datasets for testing of novel systems and suggest evaluation methods (i.e., Accuracy score, F-score). Finally, Ferreira and Vlachos~\cite{ferreira2016emergent} examine rumour detection in environments that are not related to social media. They present a dataset that comprises online articles and propose tailored features to the structure of the articles. They utilise logistic regression to categorise articles into those which are verified and those that are false with relative success (73\% accuracy).

\section{Methodology}
\label{sec:methodology}

In this work, we propose that the problem of message stance classification is more efficiently approached by semi-supervised learning algorithms. We argue that other supervised machine learning approaches, even though they may achieve marginal higher accuracy in limited datasets, they do not perform satisfactorily at large scale, which is more relevant to real-life applications. To this end, we use a class of graph-based semi-supervised algorithms, namely Label Propagation and Label Spreading, to illustrate our arguments. It is worth noting that other semi-supervised methods could be used as well, but a full comparison of such semi-supervised algorithms is beyond the scope of this study. To further motivate our proposed methodology, below we briefly discuss the pros and cons of supervised, unsupervised and semi-supervised learning approaches.

First, supervised approaches have limitations as it pertains to capturing the diversity of the messages and the stance of the same message towards two opposite rumours. A supervised approach uses a large dataset of messages for training a model. When applying this model to a new message, the supervised approach usually ignores the original claim, towards which the message takes a positive, neutral or negative position. For example, consider two rumours, the first claiming ``X is true'' and the second claiming ``Y is true''. In a supervised approach, a message saying ``X is true and Y is not true'' trained on the first rumour will always be classified in the ``supporting'' class, irrespective of whether it refers to the first or the second rumour.

There have been hybrid supervised approaches, e.g. \cite{Lukasik2015ctl}, that take into account annotated messages from the rumour under consideration in order to enhance performance. However, these approaches have a serious drawback. In a live environment, where speed is as essential as accuracy, they require the retraining of a large set of annotated messages for every new rumour. The training of accurate supervised models can be computationally very expensive and time-consuming, which makes such hybrid approaches inappropriate for real-life applications.

Unsupervised machine learning splits the messages into distinct clusters, but it provides no details about the content of these clusters. it is therefore necessary to manually inspect a sample of messages from each cluster to decide whether the cluster consists of supporting, denying or neutral messages.


This brings us to the semi-supervised learning, where only a few observations are labelled and are used as seeds for the algorithm to cluster the remaining input data correctly. This approach has several advantages.
First, it requires only a few labelled observations, therefore the end-user only has to manually tag a small number of messages.
Second, it is faster than supervised approaches, such as \cite{Lukasik2015ctl}, which require recalibration while new messages from the rumour under consideration are being collected. Finally, it is rumour-specific, i.e. it allows the same text to be classified in different classes for different rumours, depending on the content of the rumour claim.

\subsection{Data Description}
\label{sec:data-description}

Our dataset consists of the 72 rumours used in \cite{Giasemidis:2016dtv}
\footnote{This dataset is not currently publicly available due to Intellectual Property (IP) reasons. However, the method is validated on two publicly available datasets in Section \ref{sec:comparison}, where the results are reproducible.}. 
The rumours were manually identified from messages (tweets) collected from Twitter, using the Twitter public API and searching for keywords related to specific events. All messages were manually annotated as \textit{supporting}, \textit{neutral/questioning} or \textit{against} towards the corresponding rumour.

\begin{figure*}
	\centering
	\begin{subfigure}[t]{0.45\textwidth}
		\centering
		\includegraphics[width=\linewidth]{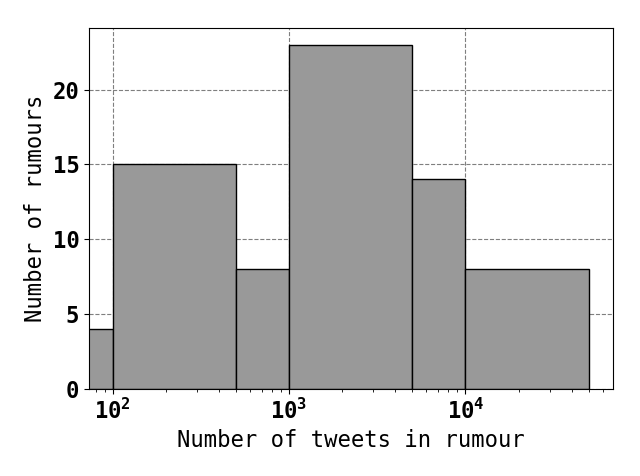}
		\caption{Total}
		\label{fig:tweet-distribution-total}
	\end{subfigure}
	~
	\begin{subfigure}[t]{0.45\textwidth}
		\centering
		\includegraphics[width=\linewidth]{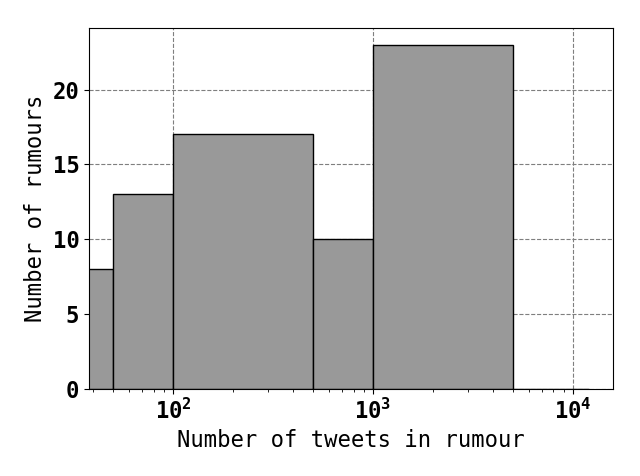}
		\caption{Original English only}
		\label{fig:tweet-distribution-original-eng}
	\end{subfigure}
	\caption{Distribution of the number of tweets in rumours: total (left) and original English only (right).}
	\label{fig:tweet-distribution}
\end{figure*}

The size of the rumours varies from 23 to 46,807 tweets, see Figure \ref{fig:tweet-distribution-total}. For tweet stance classification, only the original tweets (i.e. those that are not re-tweets) must be classified as \textit{supporting}, \textit{neutral} or \textit{against} towards the rumour, as re-tweets are assigned to the same class as their original tweet. Additionally, we skip non-English tweets, because the features we consider are language (here, English) specific.
Figure \ref{fig:tweet-distribution-original-eng} shows the distribution of the number of original English tweets for the 72 rumours.

All messages are pre-processed before feature extraction. We follow the pre-processing steps in \cite{Lukasik2015ctl}; (i) URLs, e-mails and Twitter mentions\footnote{Twitter mentions are username tags starting with the @ symbol.} are removed, (ii) text is lower-cased, (iii) all punctuations other than ``,'', ``.'', ``!'', ``?'' are removed, (iv) multiple occurrences of characters are replaced with double occurrence, and (v) extra white space is removed. Stemming, i.e. the process of reducing inflected (or sometimes derived) words to their word stem, is not performed as the Brown clusters (see Section \ref{sec:feature-space}) include whole words. Stop-words are included, because they capture important features, such as negation. 

\subsection{Feature Space}
\label{sec:feature-space}

The messages are vectorised using three different strategies (feature sets), that are proposed in the literature (and combinations thereof).
\begin{enumerate}
	\item 1,000 Brown clusters, denoted as ``BrownC'', extracted in \cite{Owoputi2013} from a Twitter corpus. Every word in a tweet is placed in one of the 1,000 clusters, which represent the features.

	\item Linguistic features, denoted as ``Ling'', such as complexity of the message, number of tentative words (e.g. ``confuse'', ``suppose'', ``wonder''), that indicate uncertainty, number of swearing words, sentiment, negation, etc. These features aim to capture statistical patterns, such as tentative words being more common in messages that question a claim.

	\item 2-grams to 6-grams features (abbreviated as ``NGrams'') of the messages in a rumour. It is worth noting that the total number of N-grams (i.e. features) varies from rumour to rumour, in contrast to the aforementioned feature sets where the features are fixed for all rumours. Another drawback is that, as new messages arrive in a live system, the feature-set is expanding. To apply this feature set, a sufficient number of messages must have been collected to capture the diversity in the N-grams.

	\item A combination of Brown clusters with the sentiment and negation features from the Linguistic features (``Brown \& Ling''). We choose these two linguistic features, as we would like to study the effect of the sentiment and negation in message stance classification performance. \footnote{Sentiment was extracted using the ``Vader Sentiment Analyser'' and negation was estimated using the ``Stanford Dependency Parser'', with the NLTK library in Python.}
\end{enumerate}
Feature selection is performed to choose the best feature set that represents the data; however we do not attempt to further reduce the size of each feature set as these are standard feature sets to represent the language in Natural Language Processing (NLP) problems \cite{yi2003sentiment}.

\subsection{Label Propagation and Label Spreading}
\label{sec:method-label-propagation}

Label propagation (LP) is a semi-supervised machine-learning method, in which observations (here messages) are represented as nodes on a graph (see \cite{Chapelle2016} for a review). Consider a graph $g = (V, E)$, where $V = \{v_1, \ldots, v_n\}$ is the set of vertices (here messages), corresponding to the data (feature vectors) $X = \{\mathbf{x}_i \in \mathbb{R}^{m}| i = 1, \ldots, n\}$, and $E$ is the set of edges, representing the similarities between the nodes, through a similarity matrix $W$. A typical choice of similarity matrix is the Gaussian kernel with width $\sigma$, i.e.
\begin{equation}
	W_{ij} = \text{e}^{-\frac{|| \mathbf{x}_i - \mathbf{x}_j ||^2}{2\sigma^2}}.
	\label{eq:similarity-matrix}
\end{equation}
The width $\sigma$ is a free parameter that requires selection. 
The graph could be fully connected or a $k$-nearest neighbours graph.

Given the graph $g$ and a subset of labelled observations, the LP algorithm aims to propagate the labels on the graph, each node propagating its label to its neighbours until convergence. 

Let $\mathbf{y}_i = (y_{i,1}, y_{i,2}, y_{i,3}) \in \mathbb{R}^3$, where $y_{i,j}$ is the probability of observation $i$ being in class $C_j \in \{-1, 0, 1\}$ representing the three classes corresponding to against, neutral and supporting messages, respectively. We denote 
$Y_l = \{\mathbf{y}_1, \ldots, \mathbf{y}_l\}$ the set of $l$ labelled observations, with typically $l << n$, where $y_{i, j} = 1$ and $y_{i, k} = 0$ for $k\neq j$. Also, let $Y_u = \{\mathbf{0}, \ldots, \mathbf{0}\}$ be the set of the $n-l$ unlabelled observations, where $\mathbf{0} \in \mathbb{R}^3$ is the null vector. The algorithm proceeds as follows:
\begin{enumerate}
	\item Compute similarity matrix $W$.
	\item Compute the diagonal degree matrix $D$, $D_{ii} = \sum_j W_{ij}$.
	\item Initialise the labels $\hat{Y}^{(0)} \leftarrow (Y_l, Y_u)$.
	\item Iterate and impose hard-clustering: $\hat{Y}^{(t+1)} \leftarrow D^{-1} W \hat{Y}^{(t)}$ and $\hat{Y}^{(t+1)}_l \leftarrow Y_l$, where $t$ is the iteration step, until convergence.
\end{enumerate}

In step 4, the algorithm assigns the average label (or probability of class membership) of the neighbours of a vertex $v_i$ to vertex $v_i$, i.e. 
\begin{equation}
	\hat{\mathbf{y}}_i^{(t+1)} = \frac{\sum_{j=1}^n W_{ij} \hat{\mathbf{y}}_j^{(t)}}{D_{ii}},
	\label{eq:labelpropagation}
\end{equation}
The vertex is assigned to the class with the highest probability, i.e. $C_i = \arg\max \mathbf{y}_i$.
The proof for convergence is beyond the scope of this study, but the interested reader should refer to \cite{Zhu2002} and \cite[Chapter 11]{Chapelle2016}.

Variations of this algorithm allow for soft clustering, i.e. permitting the labelled data to change their cluster, by removing the hard-clustering assignment in step 4. This is achieved by introducing a parameter $\alpha \in [0,1]$ in the numerator and denominator of Eq. \eqref{eq:labelpropagation} for the labelled data.

A similar algorithm, called Label Spreading (LS), uses the normalised Laplacian in the iteration step 4 above and allows the tagged observations to change classes. The algorithm becomes:
\begin{enumerate}
	\item Compute similarity matrix $W$, with $W_{ii} = 0$.
	\item Compute the diagonal degree matrix $D$, $D_{ii} = \sum_j W_{ij}$.
	\item Compute the normalised graph Laplacian $L_s = D^{-1/2} W D^{-1/2}$.
	\item Initialise the labels $\hat{Y}^{(0)} \leftarrow (Y_l, Y_u)$.
	\item Choose a parameter $\alpha \in [0, 1]$.
	\item Iterate $\hat{Y}^{(t+1)} \leftarrow \alpha L_s \hat{Y}^{(t)} + (1-\alpha)\hat{Y}^{(0)}$ until convergence.
\end{enumerate}
The algorithm has been proved to converge, see \cite{Zhou2004} and \cite[Chapter 11]{Chapelle2016} for further details.

The computational time of these algorithms is of order $\mathcal{O}(n^3)$ for dense graphs and $\mathcal{O}(n^2)$ for sparse ones \cite[Section 11.2]{Chapelle2016}.

The cost function must consider both the initial labelling and the geometry of the data induced by the graph structure (i.e. edges and weights $W$) \cite{Zhou2004,Chapelle2016},
\begin{equation}
	\sum_{i=1}^l || \hat{\mathbf{y}}_i - \mathbf{y}_i ||^2 + \frac{1}{2} \sum_{i,j=1}^n W_{ij} || \hat{y}_i - \hat{y}_j ||^2,
	\label{eq:lp-cost}
\end{equation}
where the first term is a fitting constraint for the labelled data and the second term heavily penalises points that are close in the feature space but have different labels (smoothness constraint). 


In this study, we use the algorithms as implemented in the \textit{scikit-learn} library in Python \cite{scikit-learn} with a bug fix\footnote{\url{https://github.com/scikit-learn/scikit-learn/pull/3751/files}} that allows hard-clamping for $\alpha=1$.


\section{Experimentation and Results}
\label{sec:results}

In this section, we experiment with different settings of the algorithms (such as feature sets, kernels, selection of hyper-parameters), before we validate it with two publicly available datasets. The experimentation will lead to the final model and involves the following steps:
\begin{itemize}
	\item Selection of feature set, see Section \ref{sec:feature-space}.
	\item Selection between Label Propagation and Label Spreading, see Section \ref{sec:method-label-propagation}.
	\item Selection between Gaussian and $k$-nearest neighbours kernels.
	\item Selection of the kernel's hyper-parameter $\sigma$.
\end{itemize}

\subsection{Label Propagation and Label Spreading}
\label{sec:results-label-propagation}

The Label Propagation and Label Spreading methods require the selection of a hyper-parameter, depending on the kernel used to generate the graph. 

For Gaussian (or ``rbf'') kernel, defined in eq. \eqref{eq:similarity-matrix}, and fully connected graph, this is the parameter $\sigma$. We use a grid-search for finding the optimal parameter, searching in a set of values that span different orders of magnitude of $\sigma$ from $\mathcal{O}(10^{-1})$ to $\mathcal{O}(10^3)$. As we see below, this range of values is sufficiently large in the search for the optimal parameter.

For $k$-nearest neighbours, we experiment with different numbers, $k$, of nearest neighbours when constructing the similarity matrix, from $5$ to $50$ in increments of $5$.

The semi-supervised algorithm requires a sample of annotated (manually classified) messages. For our experiments we annotate the first $N$ messages (chronologically) that appear in a rumour, where the number of  manually annotated messages is gradually increased $N = \{10, 20, 30, 40, 50\}$ for each rumour. Therefore, we skip rumours with less than 50 original tweets, resulting in a total of 64 rumours. We validate the performance of the model on each rumour using the messages that are not initially annotated. 

We compute several performance scores, such as the accuracy, the weighted accuracy, F1-score and log-loss (entropy) scores.
The accuracy is not a good performance score for biased datasets, which is the  case in tweet stance classification, as most messages are in favour of the rumour. For this reason, we focus on weighted accuracy, F1-score and entropy for choosing the best-performing feature set, kernel and hyper-parameter.

We also experiment with different features sets, pre-processing steps, kernel (e.g. $k$-nearest neighbours (``knn'')) and algorithms (e.g. Label Spreading (LS)). 
We summarise these results in Table \ref{tbl:summary-of-results}, where we show the maximum accuracy, weighted accuracy, F1-score 
for $N=50$ annotated messages. We also present the values of accuracy and F1-score at the optimal parameter (``opt param''), which is the value of the parameter ($\sigma$ or $k$) where the weighted accuracy is maximised. The BrownC$^{*}$ feature set was created by altering two pre-processing steps, i.e. (i) stemming is performed, and (ii) stop-words are removed.

\begin{table*}
	\centering
	\begin{tabular}{|l||p{1.55cm}|p{1.9cm}|p{1.55cm}|p{1.55cm}|p{1.8cm}|p{1.8cm}|}
		\hline
		Method & Max Accuracy & Max Weighted Accuracy & Max F1-Score & Accuracy at opt param & F1-Score at opt param \\ \hline \hline
		LP-BrownC-rbf			& 0.7822	& 0.4995	& 0.4606	& 0.7434	& 0.4500	\\ \hline
		LP-BrownC$^{*}$-rbf		& 0.7604	& 0.4891	& 0.4341	& 0.7068	& 0.4151	\\ \hline
		LP-Ling-rbf				& 0.7529 	& 0.4458	& 0.3997	& 0.6725	& 0.3829	\\ \hline
		LP-BrownC \& Ling-rbf	& 0.7678	& 0.4736	& 0.4318	& 0.7083	& 0.4228 	\\ \hline
		LP-Ngrams-rbf			& 0.7593	& 0.4389	& 0.3722	& 0.7001	& 0.3577	\\ \hline
		LP-BrownC-knn			& 0.7435	& 0.4112	& 0.3713	& 0.7141	& 0.3713 	\\ \hline
		LS-BrownC-rbf			& 0.793		& 0.5037	& 0.4763	& 0.7489	& 0.4666	\\ \hline
	\end{tabular}
	\caption{Summary of performance scores for different methods. The first column contains the method (algorithm-feature set-kernel). The next four columns contain the maximum scores (occurring at different parameters) for $N=50$ annotated messages. The last two columns contain the accuracy and F1-score at the optimal parameter, i.e. the parameter where the weighted accuracy is maximised.}
	\label{tbl:summary-of-results}
\end{table*}

First, we observe that using whole words and neglecting stemming together with the use of stop-words yields better results. We investigated the effect of stemming and stop-words separately (not shown in the table however). We found that either stemming or stop-word removal results in lower performance scores. In addition, we notice that linguistic and N-gram features are poor indicators for message stance classification. Combining the Brown clusters with sentiment and negation linguistic features (BrownC \& Ling) does not increase performance. This is because sentiment is also not a good indicator of message stance. 

Between the available kernels, the $k$-nearest neighbours (``knn'') appears to perform worse. We understand this to be due to the fact that the $k$-nearest neighbours kernel assigns either a hard-link (of unit weight) or no link between nodes with no weighting to capture the degree of similarity between messages.

\begin{figure}
	\centering
	\includegraphics[width=\linewidth]{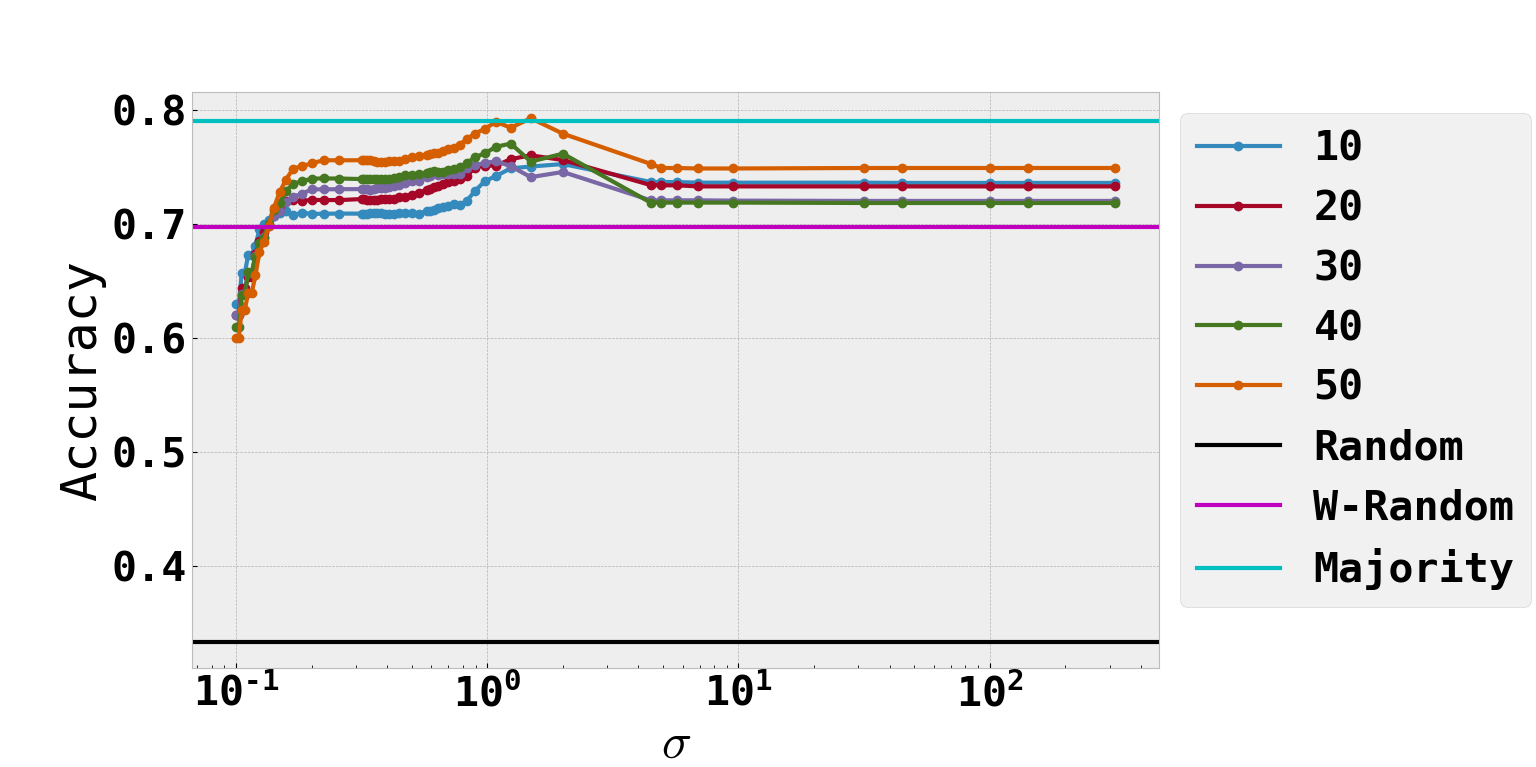}
	\caption{Accuracy of the LS algorithm with rbf kernel and Brown cluster features against $\sigma$-parameter for several numbers of annotated messages $N$.}
	\label{fig:accuracy1}
\end{figure}

\begin{figure}
	\centering
	\includegraphics[width=\linewidth]{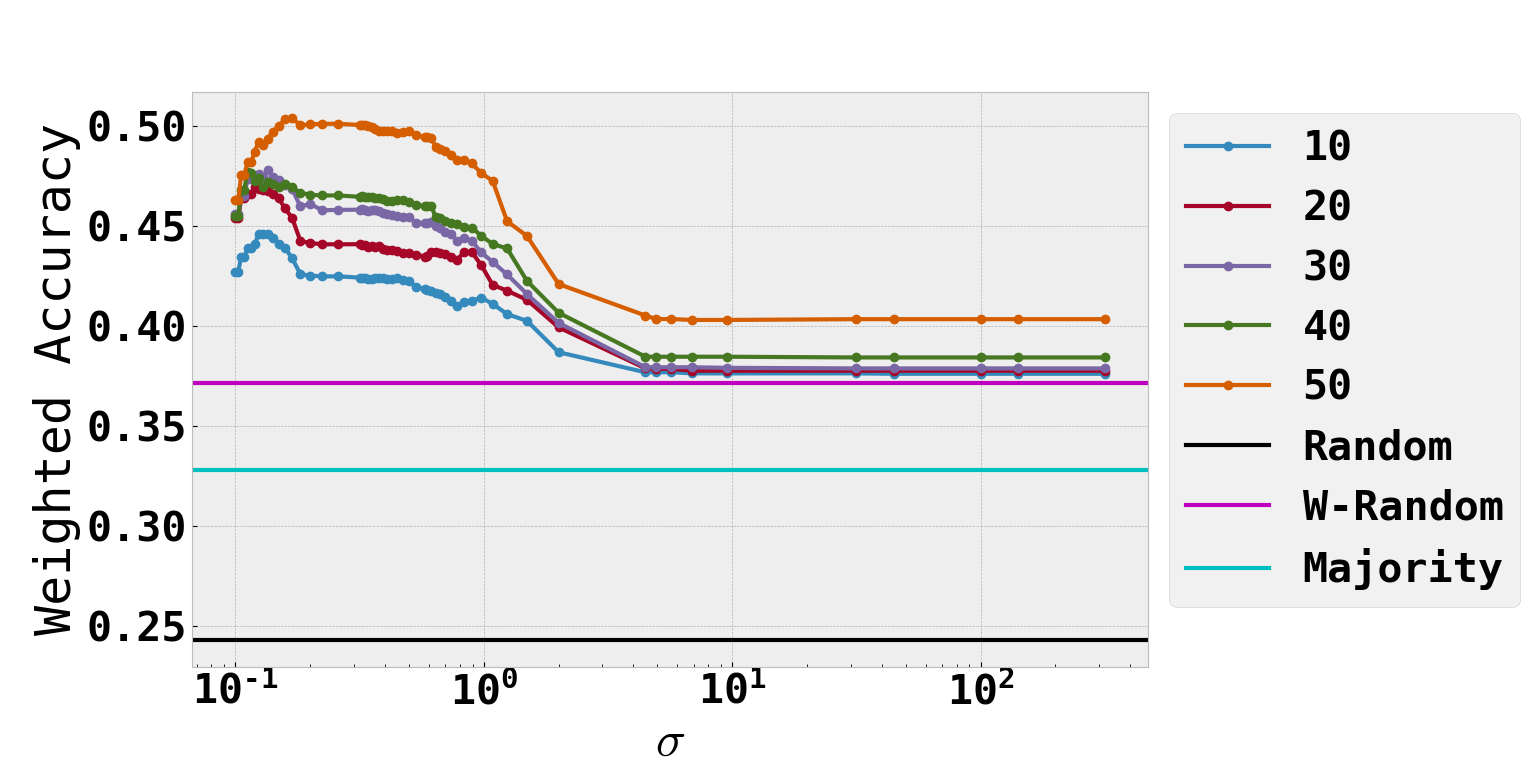}
	\caption{Weighted accuracy of the LS algorithm with rbf kernel and Brown cluster features against $\sigma$-parameter for several numbers of annotated messages $N$.}
	\label{fig:waccuracy1}
\end{figure}

\begin{figure}
	\centering
	\includegraphics[width=\linewidth]{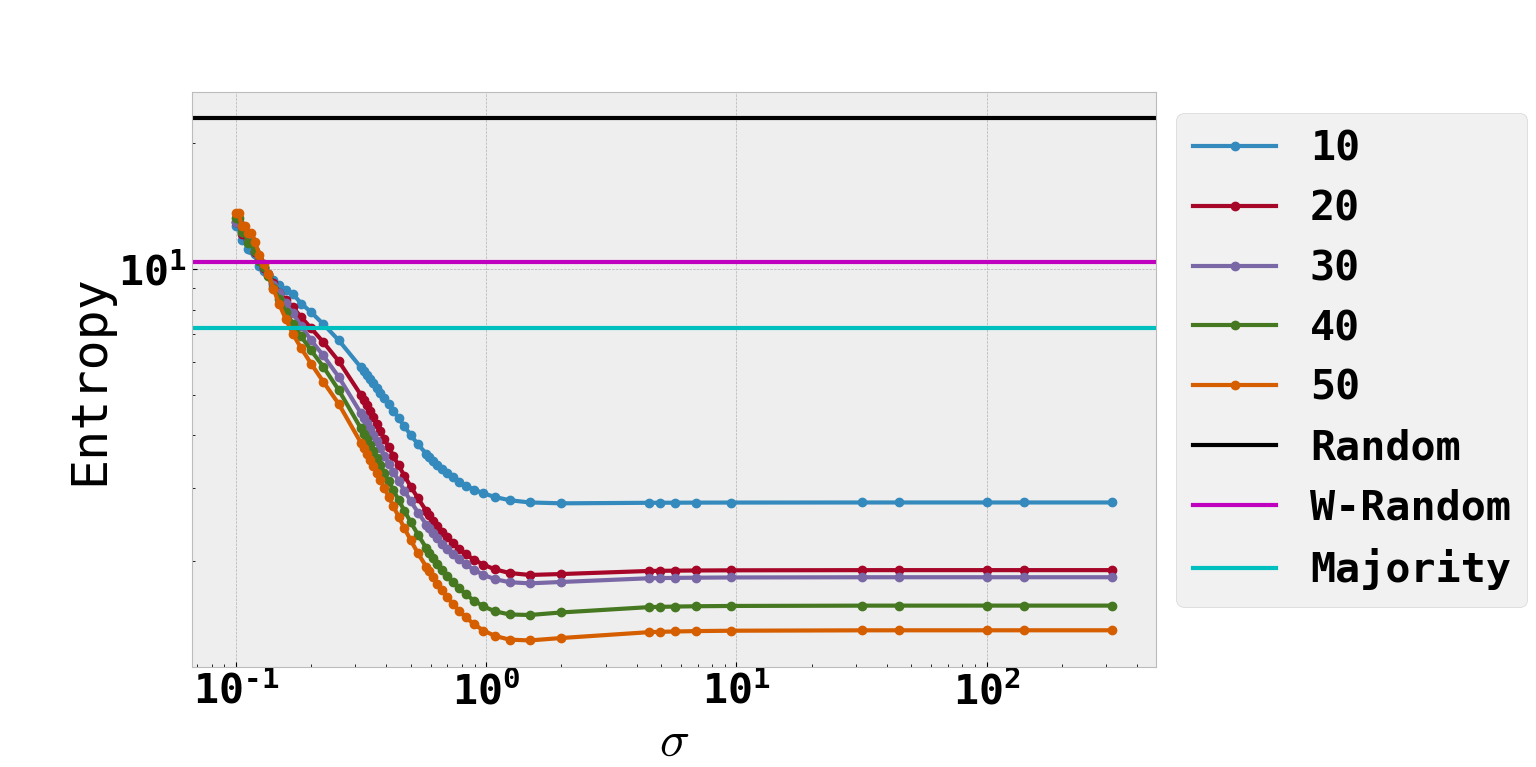}
	\caption{Entropy of the LS algorithm with rbf kernel and Brown cluster features against $\sigma$-parameter for several numbers of annotated messages $N$. The $y$-axis is in log-scale to highlight the local minimum of $\sigma$.}
	\label{fig:entropy1}
\end{figure}

Finally, the Label Spreading (LS) algorithm delivers very similar results to LP, performing marginally better. The performance plots for LP appear very similar to those in Figures \ref{fig:accuracy1}--\ref{fig:entropy1}, with the optimal scores being at the same regions of the $\sigma$-parameter. The two algorithms only differ, when $\alpha = 1$, at the normalisation of the weight matrix, $W$. Therefore, their results are expected to be very similar.

We now focus on the best-performing method (``LS'' algorithm with ``rbf'' kernel and ``BrownC'' feature set) and explore its parameter space in more detail. In Figures \ref{fig:accuracy1}--\ref{fig:entropy1}, we plot the average performance scores for the 64 rumours as a function of the $\sigma$-parameter for different numbers of annotated messages. For comparison we also plot three benchmark models: a random classifier, which randomly assigns a message to a class with probability 1/3, a weighted random classifier, which classifies a message in proportion to the class-frequency in a rumour, and the majority model, which assigns all messages to the majority class of a rumour.

We observe that the accuracy and weighted accuracy increase and the entropy decreases as the number of annotated messages increases, as expected. The more initial information the algorithm has, the better it performs. In addition, in most cases the models outperform the benchmark models \footnote{The majority model outperforms the semi-supervised models on the accuracy score for some values of the parameters, but this is an artefact of the biased dataset, which becomes evident when looking at the weighted accuracy, Figure~\ref{fig:waccuracy1}, and entropy, Figure~\ref{fig:entropy1}.}.

In more detail, we observe that all metrics have a constant plateaux for $\sigma \gtrsim 5$. These values suppress the exponent of the kernel \eqref{eq:similarity-matrix}, resulting in a similarity matrix (cf. eq. \eqref{eq:similarity-matrix}) whose elements are all very close to $1$. Therefore, all messages appear to be very similar to each other, hence giving the same prediction. For smaller values of $\sigma$, the messages become distinguishable in the graph representation, resulting in an increase of accuracy and weighted accuracy, where the entropy has a local minimum. For very small values of $\sigma$, the exponent in \eqref{eq:similarity-matrix} becomes too large, hence, the elements of the similarity matrix become too small and the messages are very weakly connected in the graph representation, resulting in poor performance. It is the region at $\sigma \sim \mathcal{O}(1)$, where the accuracy scores have a local maximum and the entropy has a local minimum.

Focusing on the accuracy and entropy, the local optimal occurs at value $\sigma \sim 0.85$, whereas the weighted accuracy shows a fluctuating plateaux for $0.2 < \sigma \lesssim 1 $. Combining the conclusions from the three metrics, we choose $\sigma = 0.85$ as the optimal value.

The remaining methods considered in Table \ref{tbl:summary-of-results} behave similarly, showing the same qualitative patterns, although the location of the optimal parameter may differ.

\begin{figure}
	\centering
	\includegraphics[width=\linewidth]{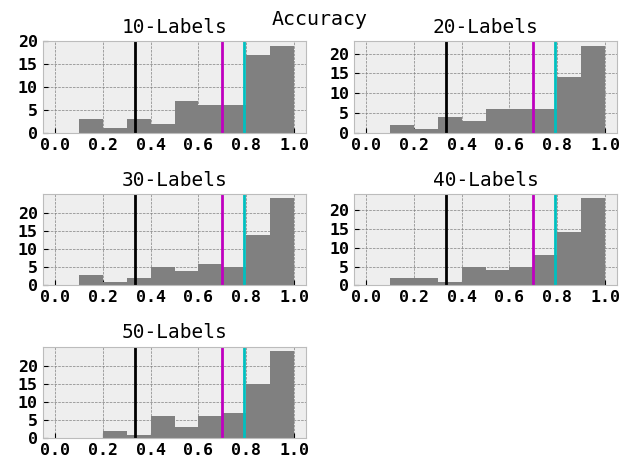}
	\caption{Distribution of rumour accuracies for increasing number of annotated messages. The vertical lines indicate the accuracy of the benchmark models, random (black), weighted random (magenta) and majority (cyan).}
	\label{fig:accuracy-distribution}
\end{figure}

In Figure \ref{fig:accuracy-distribution}, we plot the distribution of the accuracies of the 64 rumours for the LS algorithm with rbf kernel and $\sigma=0.85$. We notice that as more messages get annotated, the distribution is shifted to higher values. For $N=50$, more than half of the rumours have accuracy greater than 80\%. Only two rumours show an accuracy less than random, which will be investigated in future work, see Section \ref{sec:conclusion}. Some rumours have low accuracy because one or two classes are not present in the first 50 annotated messages. We aim to resolve such cases in future work, see also the discussion in Section \ref{sec:conclusion}.

Here, we selected $\sigma$ so that it optimises the average performance scores. However, the messages in different rumours might have distinct spread in the feature space, hence, requiring a varying $\sigma$ that depends on the particular rumour. 

In \cite{Zhu2002}, the authors proposed a heuristic method for determining the $\sigma$ of individual datasets (here rumours). Particularly, they find the minimum spanning tree of labelled data, from which they estimate the minimum distance between two nodes that belong on different classes. Then $\sigma$ is set to one third of that distance, following the rule of $3\sigma$ of the normal distribution. 

\begin{figure}
	\centering
	\begin{subfigure}[t]{0.47\textwidth}
		\centering
		\includegraphics[width=\textwidth]{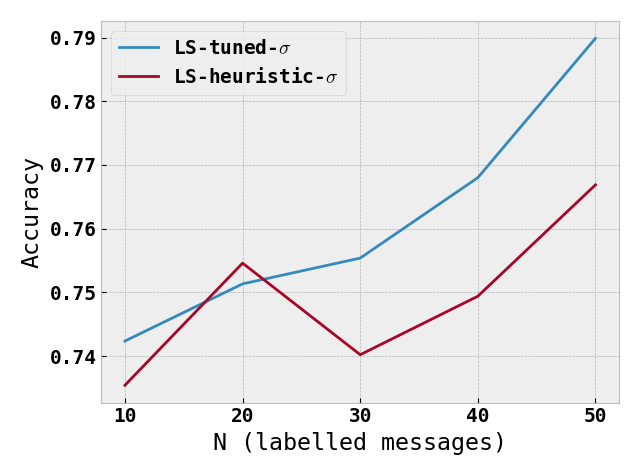}
		\caption{Accuracy}
		\label{fig:accuracy-tuned-heuristic}
	\end{subfigure}
	\begin{subfigure}[t]{0.47\textwidth}
		\centering
		\includegraphics[width=\textwidth]{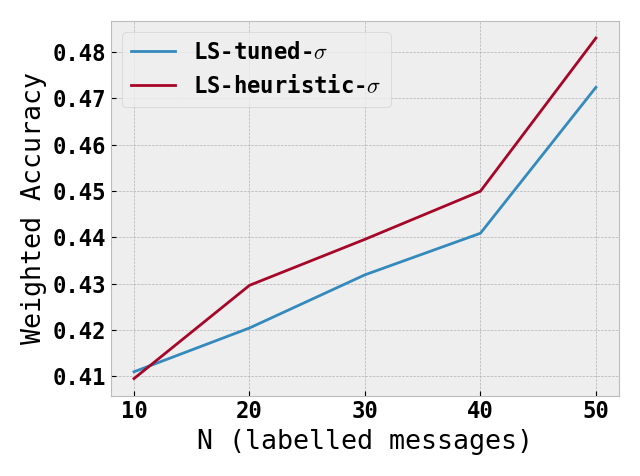}
		\caption{Weighted accuracy}
		\label{fig:waccuracy-tuned-heuristic}
	\end{subfigure}
	
	\caption{Average performance of the LS method with tuned and heuristic method for finding $\sigma$}
	\label{fig:LS-tuned-heuristic}
\end{figure}

In Figure \ref{fig:LS-tuned-heuristic}, we plot the performance scores of the LS with tuned $\sigma=0.85$ and the LS with $\sigma$ dynamically determined using the heuristic of \cite{Zhu2002}. We observe that the accuracy of the ``tuned'' method is higher than that of the ``heuristic'' method; however, the latter outperforms the former in the weighted accuracy, indicating that the ``heuristic'' $\sigma$ method is better for biased datasets. 

Although the ``heuristic'' method underperforms in terms of accuracy, it is sometimes useful in a real-world system, which operates on corpus other than Twitter. 

\subsection{Validation and Comparison}
\label{sec:comparison}

We validate the proposed algorithm on two datasets in \cite{Lukasik2015ctl} and \cite{zubiagaSemEval2016} and compare the performance of our approach with the Gaussian Processes of \cite{Lukasik2015ctl}. We choose to compare against this study for the following two reasons. First, its dataset and algorithms are publicly available. 
This allows us to make a direct comparison on the same dataset. Second, the authors of this work use a hybrid approach, which, to the best of our knowledge, is among the state-of-the-art in the academic literature.

The authors of \cite{Lukasik2015ctl} considered Gaussian Processes in three different training methods. The first method (here denoted as ``GP'') involves training only on the first $N$ annotated messages from the rumour under consideration (the target rumour). In the second method (``GPPooled''), a GP model is trained on messages from other rumours in the dataset (the reference rumours) combined with the first $N$ messages from the target rumour. The final configuration (``GPICM'') is similar to the second one, but instead weighs the influence from the reference rumours.

We focus on the Brown clusters excluding the bag-of-words features. Our methods consist of the LS algorithm with rbf kernel and $\sigma$ either tuned to $\sigma=0.85$ or determined by the heuristic approach, described in the previous section, for each rumour.

\subsubsection{London Riots Dataset}
\label{sec:london-riots-dataset}

The dataset in \cite{Lukasik2015ctl} consists of seven rumours from the London riots in 2011. Due to anonymisation of the dataset, messages are replaced with their features, i.e. the 1,000 Brown clusters and bag of words. Therefore, we perform no pre-processing of the messages and work directly with their feature representation. 

In \cite{Lukasik2015ctl}, the authors trained a Gaussian Process (GP) \footnote{For an introductory review on GP see \cite{Roberts20110550}.} using only original tweets and validated it on a set that included both original tweets and retweets. Similarly, we annotate the first $N = \{10, 20, 30, 40, 50\}$ original tweets and compute the performance scores using all the remaining tweets. Here, we are not able to simply assign every retweet to the same class as its original tweet because the dataset has no retweet id information, from which one can associate retweets to original tweets. Instead, the dataset includes a tag identifying whether the message is a retweet or not. Therefore, the retweets participate in the algorithm as ``original tweets".

\begin{figure}
	\centering
	\begin{subfigure}[t]{\linewidth}
		\centering
		\includegraphics[width=\linewidth]{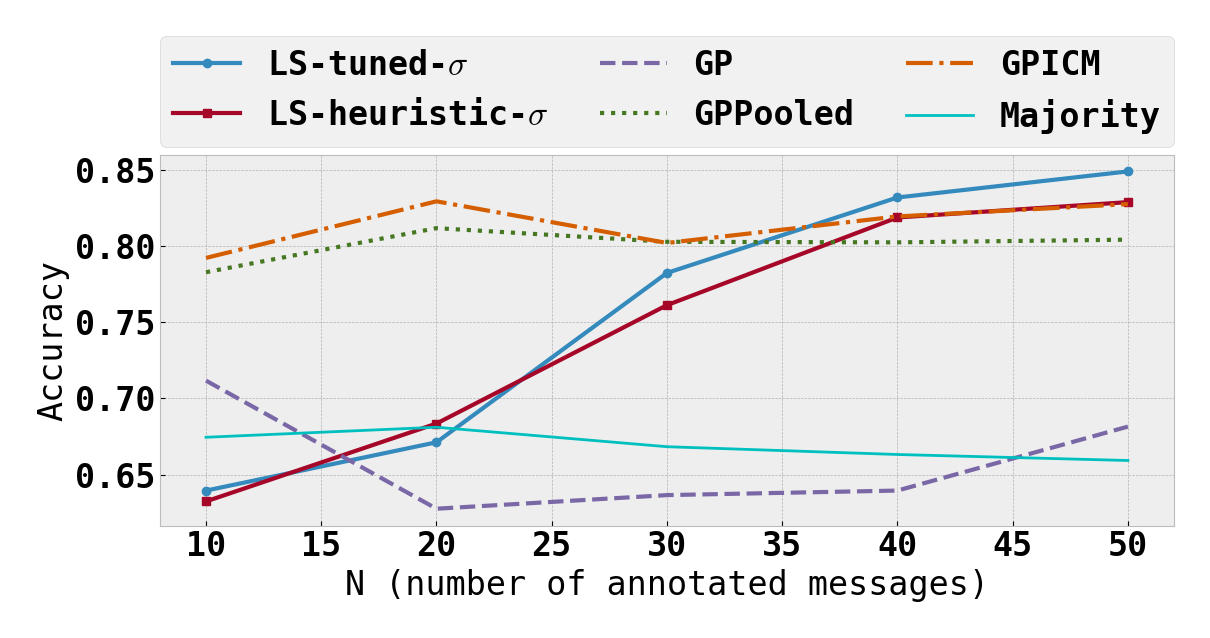}
		\caption{Accuracy}
		\label{fig:accuracy-validation}
	\end{subfigure}
	\begin{subfigure}[t]{\linewidth}
		\centering
		\includegraphics[width=\linewidth]{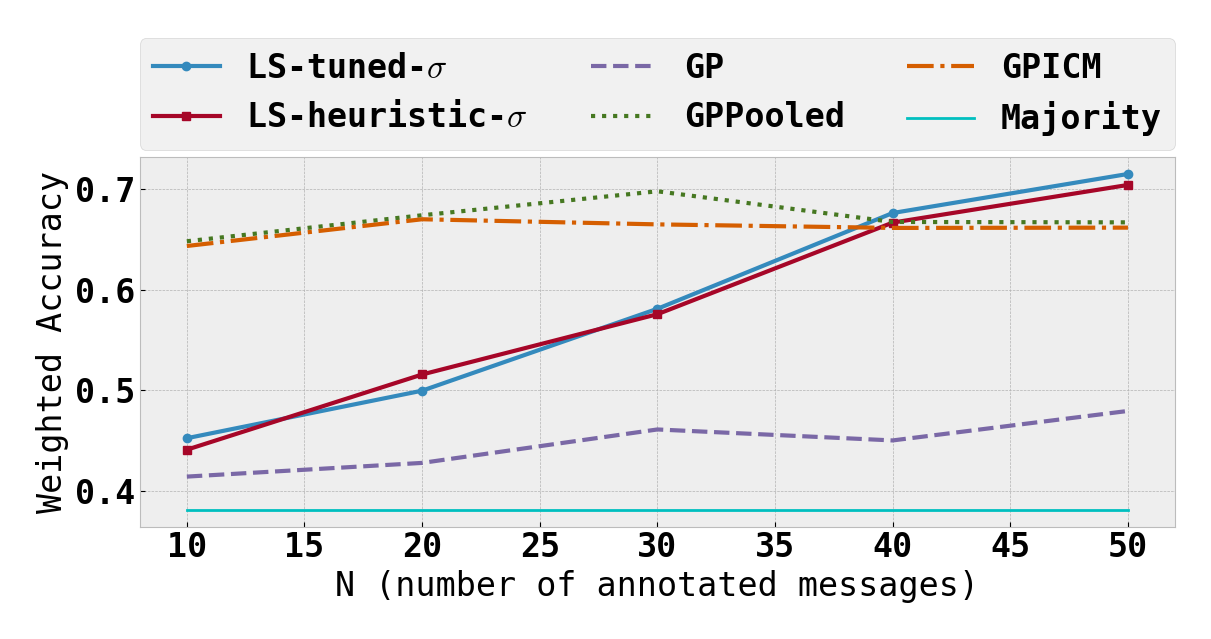}
		\caption{Weighted accuracy}
		\label{fig:waccuracy-validation}
	\end{subfigure}
%

	\caption{Average performance scores of the LS (solid line with round markers), the three GP methods (non-solid lines) and benchmark models (solid lines) of the 7 London riots rumours.}
	\label{fig:scores-validation}
\end{figure}

The accuracy and weighted accuracy of the two proposed semi-supervised methods as a function of $N$ are plotted in Figure \ref{fig:scores-validation}. For comparison, we also plot the performance scores of the three GP methods and benchmark models. Regarding the LS algorithm, we observe that the performance scores increase as more tweets get annotated. Particularly, the tuned $\sigma$ method achieves an accuracy of $83.2\%$ and $84.9\%$, whereas the ``heuristic'' $\sigma$ method scores $81.9\%$ and $82.9\%$, at $N=40$ and $N=50$ respectively. All performance scores show that the LS method outperforms all other methods for $N\geq40$. Particularly, it outperforms the ``GP'' method, which is actually a semi-supervised approach for Gaussian Processes, for all $N$. Although the remaining two GP methods achieve higher performance at early stages, they suffer from scalability and speed issues, hence, they are inefficient for quick message stance classification in a rapid-response live system.

For example, when applied on this dataset, consisting of 7 rumours and $7297$ tweets (which is a moderate number, for real-life situations), the GP methods required about a week of training, on a $12$-core machine \footnote{Dual Intel(R) Xeon(R) CPU E5-2630 v2 @ 2.60GHz, 256GB RAM.}. Given that frequent retraining would be required for any live system, this demonstrates that supervised methods, though accurate, cannot scale up, therefore limiting their usefulness for realistic systems.

\subsubsection{PHEME Dataset}
\label{sec:pheme-dataset}

We also compare the two methods on another publicly available dataset \cite{zubiagaSemEval2016}. This set consists of tweet conversations, collected in association with 9 breaking news stories. The conversations are organised in threads the root of which is the initiating rumour tweet, accompanied by the corresponding replies. The tweets are annotated for support, certainty and evidentiality. In order to align this dataset with the purpose of this study, we group threads by rumour. In our nomenclature there are two levels of support in this set; whether the initial tweet supports or not the rumour and whether the subsequent tweets support the initial tweet's claim. We straighten this two-step relation, by resolving the support of each tweet against the rumour, and update the annotation accordingly. For example if the initial tweet supports the rumour claim it is annotated as such. If a subsequent tweet (reply) negates the initial tweet with certainty, then it is annotated as not supporting the rumour claim.

The dataset contains $297$ threads containing $4561$ tweets (including retweets), spanning $138$ rumours organised in $9$ stories. For the purpose of this study, as explained in previous sections, we select only the rumours containing at least $50$ English original tweets. The final number of rumours we are using from this dataset is therefore $23$ containing $2233$ (original English) tweets.

\begin{figure} 
	\centering
	\begin{subfigure}[t]{\linewidth}
		\centering
		\includegraphics[width=\linewidth]{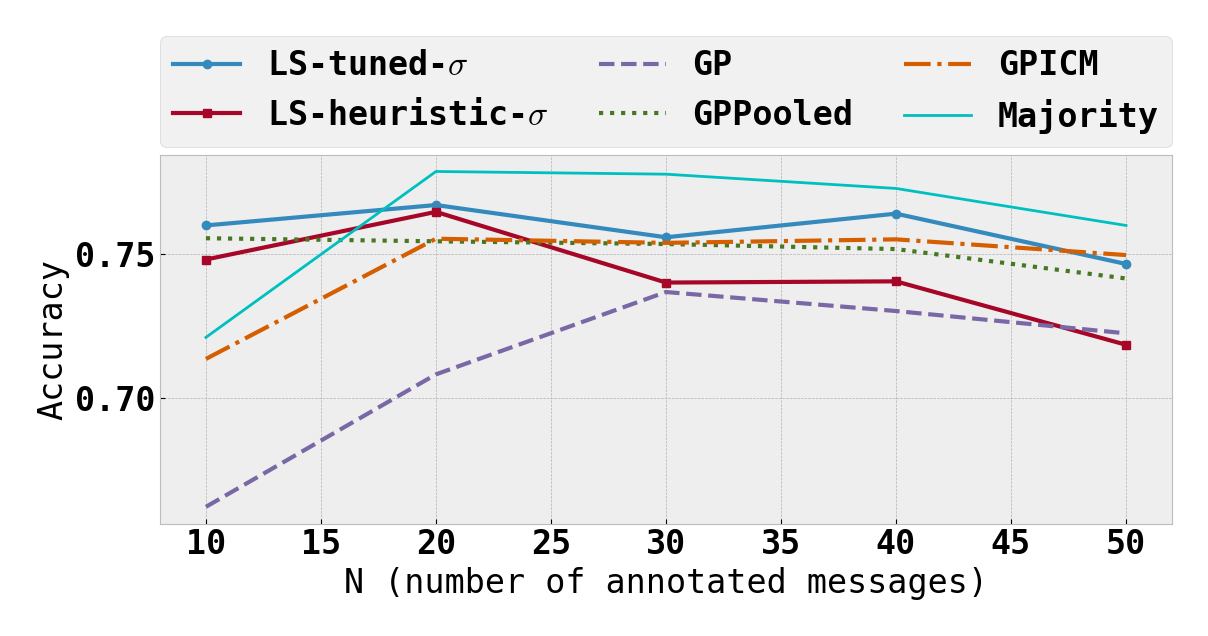}
		\caption{Accuracy}
		\label{fig:accuracy-semeval}
	\end{subfigure}
	\begin{subfigure}[t]{\linewidth}
		\centering
		\includegraphics[width=\linewidth]{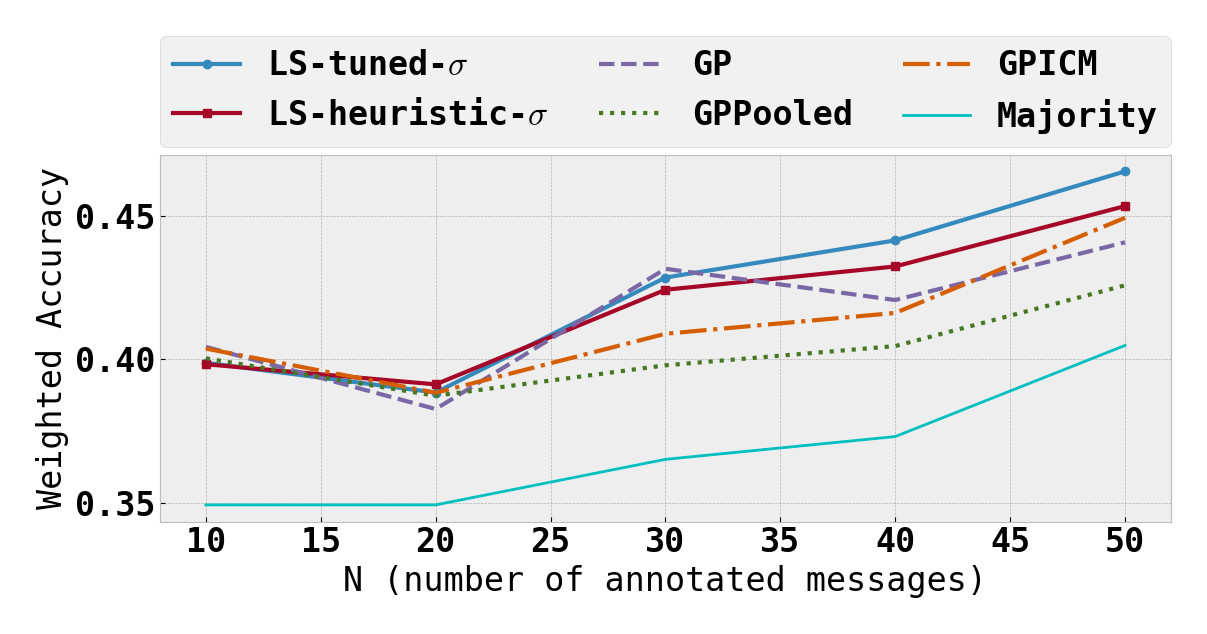}
		\caption{Weighted accuracy}
		\label{fig:waccuracy-semeval}
	\end{subfigure}
	
	\caption{Average performance scores of the LS (solid line with round markers), the three GP methods (non-solid lines) and benchmark models (solid lines) of the 23 PHEME rumours.}
	\label{fig:scores-semeval}
\end{figure}

The accuracy and weighted accuracy of the models are presented in Figure \ref{fig:scores-semeval}. Looking at the accuracy of the majority model in Figure \ref{fig:accuracy-semeval}, we conclude that the dataset is strongly biased, with most messages belonging in a single class. Therefore, we focus on the weighted accuracy, which suppress the majority model as well as the ``GPPooled'' model. Both versions of the semi-supervised LS methods, score at least as well as the GPs from the very early stages of the rumours' development and outperform the GP methods for $N \geq 40$ labelled messages. 

\subsubsection{Remarks}
Following the conclusions in \cite{Lukasik2015ctl}, we make the following observations.
\begin{itemize}
	\item The performance of LS increases as more tweets get annotated.  This behaviour is expected because a semi-supervised algorithm relies on limited information, the more the better. 

	\item ``GP'' resembles the semi-supervised learning, as only a limited number of tweets from the target rumour are used for training. Comparing to the results presented here, we achieve at least a similar accuracy from early on, $N=10$, on both datasets, however, the performance of the LS methods increases with $N$, exceeding $80\%$ at $N=40$ on the ``London Riots'' dataset.

	\item The weighted accuracy of the proposed models at $N\geq 40$ exceeds the accuracy of all three methods in \cite{Lukasik2015ctl} on both data-sets.
	
	\item ``GPICM'' in \cite{Lukasik2015ctl} outperforms the LS method for $N<30$ on the ``London Riots'' dataset (but not on the ``Pheme'' dataset). However, this might be an artefact due to the lack of messages' diversity in the ``London Riots'' dataset. GP was trained only on messages about a particular topic, the London riots, hence, all messages in the pooled rumours are relevant to the messages in the target rumour, achieving a better score due to over-fitting. This might be the reason why the GP methods do not perform significantly better at low $N$ on the second dataset, where messages from diverse topics exist.

	\item ``GPICM'' and ``GPPooled'' are particularly inefficient in a live system where both speed and accuracy are essential. Training a new model as new messages arrive slows down the process, particularly, when supervised training is performed on a large dataset, as in ``GPPooled'' and ``GPICM''.

	\item Finally, it should be noted that the ``GP'' method with no tweets from the rumour under consideration (``Leave-one-out''), simulates the situation when a completely unknown rumour is examined. Such a design could address our concerns about scalability if it performed well enough, since one would only need to train a model once. This set-up was examined but consistently underperformed every other result reported here, which is why it is not included in the plots.
\end{itemize}

Overall, the proposed algorithm achieves a better performance and is much faster than a GP (as considered in \cite{Lukasik2015ctl}). The LS scales as $\mathcal{O}(n^2)$. Specifically, the times required to process the rumours in our dataset fit the polynomial $\text{time} (n) = 2.06 \cdot 10^{-7} n^2 + 4.47 \cdot 10^{-5} n - 9.32 \cdot 10^{-3}$ seconds, i.e. a rumour with 1,000 messages is processed in 0.24 seconds \footnote{On a laptop with 16GB RAM and Intel(R) Core i7-3610QM CPU @ 2.30GHz}.

In the previous comparison we have focused on the accuracy (measured with three different metrics) of each algorithm. Here, we would like to emphasize another point of comparison, namely scaling. The top-performing Gaussian Process algorithms, i.e. ``GPPooled'' and ``GPICM'' rely on a sizeable reference library of messages, over which training is performed. In a real-life system, dealing with millions of messages and hundreds of wildly diverse rumours, this reliance cripples performance. One would first have to train on this reference library, and then apply the resulting model on arriving messages. Moreover, as the messages that do not belong to the reference library grow in number, periodically retraining will become necessary, now on an even larger library. In other words, when one considers the complexity of Gaussian Process algorithms, which is $\mathcal{O}(n^3)$ or at best $\mathcal{O}(n^2)$ \cite{Roberts20110550}, one needs to remember that, in these cases, $n$ refers to the number of reference messages. Contrary to that, ``LS'', which scales as $\mathcal{O}(n^2)$, only involves the messages of the rumour under investigation and is completely agnostic to other rumours, thus $n$ is a significantly smaller number. Furthermore, since each incoming rumour is treated independently there is no training stage and no need for retraining. It therefore becomes clear, that ``LS'' is significantly better at performing in realistic environments. 
Practically, as already mentioned, the GP methods when applied on the PHEME data set, consisting of $2233$ messages (which is a moderate number, for real-life situations) required almost $14$ days of training, on a $12$-core machine. Given that frequent retraining would be required for any live system, this demonstrates that supervised methods, though accurate, cannot scale up, therefore limiting their usefulness for realistic systems.
In contrast, the LS method would take 1.1 seconds for a rumour of size $2233$.

\section{Conclusion and Outlook}
\label{sec:conclusion}

In the modern world dominated by social media interactions, unverified  stories can spread quickly, having a huge impact on people's life, particularly on situations of crisis, such as terrorist attacks, natural disaster, accidents, or even a financial impact. Determining the trustworthiness of information is a challenging and open problem. Particularly, on situations of crisis, the veracity of rumours must be resolved as quickly as possible. Therefore, speed and classification performance are equally important. Several methods have been proposed to automate the identification of rumour veracity. Towards this goal, the classification of messages, i.e. whether they support, deny or question a rumour, is a crucial feature \cite{Giasemidis:2016dtv}, as people tend to correctly judge a situation collectively (``wisdom of the crowd''). However, this task often requires manual effort. The aim is to automate this process as much as possible and reduce the burden on end-users, i.e. a fast process that requires minimum input information from an end-user, classifies whether the messages support, deny or are neutral to the rumour and feeds the classification to the rumour veracity assessor.

Having reviewed the literature, we found that most methods for message stance classification rely on supervised machine learning \cite{zubiaga2017detection}. We argued why such algorithms do not address the problem satisfactorily, in terms of accuracy, computational speed and scalability, and instead we propose that existing semi-supervised algorithms tackle the problem more efficiently, especially when dealing with large and diverse datasets. We focus on a family of graph-based semi-supervised algorithms, the Label Propagation and Label Spreading. The algorithms' accuracy increases as more messages get annotated. In a real scenario, a software tool, with a user-tailored interface, can display the first tens of messages, which are able to be annotated by the end-user in a very short time. Our study shows that the proposed algorithms are fast and accurate, exceeding 80\% average accuracy.

We compared to Gaussian processes used in a supervised and semi-supervised setting. The results show that the graph-based algorithms are faster and at least as accurate, particularly as more annotated messages become available, therefore they are more effective for implementation in a rapid-response software system.

Despite their success, these algorithms face a few challenges, some of which have been addressed, while others require further improvements, which will be explored in future work. We briefly mention these issues below.

First, from a usability point of view, the semi-supervised method proposed in this work requires a set of annotated messages. In a live-system, an analyst or end-user might urgently need an estimate of the message stance and hence of the rumour veracity. For such scenarios, we have developed a supervised logistic regression model, trained on a subset of our dataset, which can be applied to the new messages. This model captures average message characteristics, which are unrelated to the specific rumour, such as, messages that have the word ``believe" without negation are more likely to support a statement. This is a simple solution that address the ``cold-start'' problem when no annotated messages are available. Other supervised models available in the literature could be equally applied, hence integrating multiple approaches into one tool. 
We intend to address this issue systematically in future work.

Another solution to this problem could be online learning algorithms \cite{ShalevShwartz2014,Rakhlin:2016ato}, which aim to update a model as sequences of data become available and are faster and more efficient than batch-learning supervised algorithms. However, such an approach has not been developed within the context of message stance classification, hence a complete study, end-to-end implementation and its comparison to semi-supervised methods are left for future work.

A second issue regarding the the LP and LS is that the number of classes is implied by the annotated messages. For example, if the first $N$ messages belong only to two classes, then all other messages will be classified into one of these two classes. In future work, we aim to improve the algorithm, so that if a message is too distinct from the annotated ones, then it gets classified into a new cluster.

\section*{Acknowledgements}
This work was partly supported by the UK Defence Science and Technology Laboratory under Centre for Defence Enterprise grant DSTLX-1000107083. We thank Colin Singleton, Chris Willis and Nicholas Walton for their helpful comments during the project. We would also like to thank Dr. Matthew Edgington and Alan Pilgrim for their assistance in annotating part of the dataset.

\bibliographystyle{elsarticle-num}
\bibliography{bibfile}

\end{document}